\documentclass[aps,prd,preprint,superscriptaddress,showpacs]{revtex4}
\usepackage{graphicx}
\usepackage{verbatim}
\usepackage{ulem}
\usepackage{color}
\definecolor{My_red}        {cmyk}{0.00,1.00,1.00,0.20}

\newcommand{\bmat}{\left(\begin{array}}
\newcommand{\emat}{\end{array}\right)}
\newcommand{\beq}{\begin{equation}}
\newcommand{\eeq}{\end{equation}}
\newcommand{\wt}{\widetilde}

\def\ra{\rightarrow}

\def\ld{\lambda}
\def\f{\frac}

\def\bwt{\begin{widetext}}
\def\ewt{\end{widetext}}
\def\be{\begin{equation}}
\def\ee{\end{equation}}
\def\bea{\begin{align}}
\def\eea{\end{align}}
\def\bean{\begin{align*}}
\def\eean{\end{align*}}
\def\bary{\begin{array}}
\def\eary{\end{array}}
\def\bit{\begin{itemize}}
\def\eit{\end{itemize}}

\def\ra{\rightarrow}

\def\ld{\lambda}

\def\su5u1{SU(5) \times U(1)}
\def\fsu5u1{SU(5) \times U(1)'}
\def\so10{SO(10)}
\def\sq20{SO(10) \times SO(10)}

\def\ra{\rightarrow}

\def\ld{\lambda}
\def\f{\frac}

\def\L{\left(}
\def\R{\right)}

\def\ra{\rightarrow}

\def\ld{\lambda}

\def\su5u1{SU(5) \times U(1)}
\def\fsu5u1{SU(5) \times U(1)'}
\def\so10{SO(10)}
\def\sq20{SO(10) \times SO(10)}

\usepackage[centertags]{amsmath}
\usepackage{amssymb}

\begin{document}

\title{
Decaying Asymmetric Dark Matter Relaxes\\ the AMS-Fermi Tension}

\author{Lei Feng}
\affiliation{Key Laboratory of
Dark Matter and Space Astronomy, Purple Mountain Observatory,
Chinese Academy of Sciences, Nanjing 210008, China}

\author{Zhaofeng Kang}
\email{zhaofengkang@gmail.com} \affiliation{Center for High-Energy
Physics, Peking University, Beijing, 100871, P. R. China}

\date{\today}

\begin{abstract}

The first result of AMS-02 confirms the positron fraction excess
observed by PAMELA, but the spectrum is somewhat softer than that of
PAMELA. In the dark matter (DM) interpretation it brings a tension
between AMS-02 and Fermi-LAT, which reported an excess of the
electron plus positron flux. In this work we point out that the
asymmetric cosmic ray from asymmetric dark matter (ADM) decay
relaxes the tension. It is found that in the case of two-body decay
a bosonic ADM around 2.4 TeV and decaying into $\mu^-\tau^+$ can
significantly improve the fits. Based on the $R-$parity-violating
supersymmetry with operators $LLE^c$, we propose a minimal model  to
realize that ADM. The model introduces only a pair of singlets
$(X,\bar X)$ with a tiny coupling $LH_uX$, which makes the ADM share
the lepton asymmetry and decay into $\mu^-\tau^+$ along the operator
$LLE^c$.

\end{abstract}

\pacs{12.60.Jv, 14.70.Pw, 95.35.+d}

\maketitle

\section{Introduction and motivations}

Dark matter (DM) is commonly accepted as a major component of our
Universe in the present era, and its fraction in the total energy
budget is precisely determined to be 26$\%$~\cite{Planck}.
Nevertheless, the most confirmative evidences for its existence come
from its gravitational effects, which renders its particle
properties barely known. In the galaxy, DM may annihilate or decay
into the cosmic ray (CR) components like $e^{\pm}$ and $p/\bar p$,
which induces the CR excesses (or CR anomalies). Observations of
such anomalies by means of indirect DM detections can be regarded as
a smoking-gun for DM, and they may convey important information
about the particle properties of DM.

Recently, such smoking-guns have been triggered. In the year 2008,
PAMELA reported that, within the region from 10 GeV to 100
GeV~\cite{PAMELA}, the positron fraction of CR shows a sharp rising
excess over the background. This excess was extended to the higher
energy by PAEMLA and FERMI~\cite{PF}, in spite of a larger error
bar. It is of great interest to attribute the excess to extra
$e^+/e^-$ from DM decay or annihilation. More interestingly, this is
well consistent with another important excess from the 2009
Fermi-LAT, which precisely measured the total electron plus positron
flux and observed a flat excess up to the TeV
region~\cite{Fermi}~\footnote{ATIC and HESS~\cite{Hess}, etc., also
reported total flux excess, but for simplicity we only mention
Fermi-LAT hereafter.}. On the other hand, both the PAMELA
anti-proton~\cite{pbar} and Fermi-LAT diffuse gamma
ray~\cite{Ackermann,Decaying,Panci} data are well fitted by the pure
backgrounds, which then stringently restrict the possible dark
matter explanations. In practice, the inverse Campton scattering
(ICS) process of the injected $e^+/e^-$ always produces an
associated diffuse gamma ray spectrum. In the annihilating and
decaying DM scenarios, the injected $e^+/e^-$ fluxes are
respectively proportion to $\rho(r)^2$ and $\rho(r)$ with $\rho(r)$
the DM distribution function, so the decaying DM scenario is
strongly favored~\cite{Decaying} to evade the exclusion from
Fermi-LAT. In supersymmetric standard models (SSMs), the required
extremely long lifetime $\sim 10^{26}$s and leptonic final states,
can be realized by introducing proper $R-$parity violating
operators~\cite{Cotta:2010ej}, or dimension-six operators suppressed
by the GUT-scale~\cite{Arvanitaki,Kang} and further aided by some
flavor symmetry~\cite{Kang}.

Very recently, the long-expected AMS-02 experiment released the
first result of positron fraction measurement~\cite{AMS}. It
measures the fraction over a much wider energy region, from 0.5 to
350 GeV, with an unprecedented high precision. It confirms the
excess, and largely speaking, is consistent with the result of
PAMELA. However, at the same time it brings a clear tension with the
Fermi-LAT total flux
measurement~\cite{Feng:2013zca,DeSimone:2013fia,Yuan,Jin:2013nta,Hooper}.
The tension is ascribed to the fact that the spectrum of AMS-02 is
relatively softer than that of PAMELA, especially in the higher
energy region (As stressed by Ref.~\cite{AMS}, the slope of the
spectrum decreases one magnitude of order from 10 to 250 GeV.). As a
consequence, as we fit Fermi-LAT using a relatively heavy DM, the
resulted positron fraction in the higher energy region will be too
hard to fit AMS-02 (The same problem may also be encountered in the
pulsar explanation~\cite{pulsar,Hooper}). To reconcile AMS-02 and
Fermi-LAT, we may need to consider nonconventional scenarios of CR,
which can give a relatively harder $e^-$ spectrum than the $e^+$
spectrum. This is explored in the astrophysical context like
hardening the primary $e^-$ spectrum in the higher energy
region~\cite{Feng:2013zca,Yuan:2013eba,Hooper}.

Viewing from particle physics, hardening the $e^-$ spectrum can be
naturally achieved by updating the decaying DM to the asymmetric
decaying DM. In this framework, when the asymmetric DM decay
produces asymmetric final states, says $\mu^-\tau^+$ without the
corresponding conjugate state $\mu^+\tau^-$, then the asymmetric
cosmic ray with a harder $e^-$ spectrum (than the $e^+$ spectrum) is
produced. Actually, such a scenario has already been studied even at
the time when only the PAMELA data is
available~\cite{Frandsen:2010mr,Chang:2011xn}. AMS-02 result may
favor it~\footnote{A similar point was taken in the recent
work~\cite{Masina:2013yea} which appeared on the arXiv during the
preparation of this work. Ref.~\cite{Jin:2013nta} also considered
the asymmetric CR scenario but drew a negative conclusion.
Basically, that may be due to the fact that authors of
Ref.~\cite{Jin:2013nta} actually took a constant instead of
energy-dependent asymmetry.}.

This paper is organized as follows. In Section II, a model
independent fit of the AMS-02 and Fermi-LAT data is employed,
assuming that the (scalar) asymmetric DM asymmetrically decays into
a pair of charged leptons. In Section III, we propose a minimal
supersymmetric decaying asymmetric DM model to implement the idea.
The Section IV is the discussion and conclusion. A model for
$R-$parity violation is given in Appendices A.

\section{Asymmetric cosmic ray Relaxes the AMS-Fermi tension}\label{fit}

Recently, the AMS-02 collaboration reported the first result, which
confirms the excess of positron fraction observed several years ago
by PAMELA. But the measured spectrum is mildly softer than that of
PAMELA, as causes the difficulty in fitting AMS-02 and Fermi-LAT
simultaneously. A way out of this difficulty is going beyond the
symmetric CR and assuming that in the injected CR the $e^-$ fraction
is harder than the $e^+$ fraction, namely
\begin{eqnarray}\label{}
r^+(E)\equiv\f{\Phi^+(E)}{\Phi^+(E)+\Phi^-(E)}> 0.5,
\end{eqnarray}
with $\Phi^{\pm}(E)$ the flux of $e^{\pm}$, respectively. Note that
such a solution thanks to the fact that Fermi-LAT is bind to the
sign of charges. Asymmetric DM decay potentially produces such kind
of asymmetry, and in this work we consider the following three
asymmetric modes from a scalar DM $X$ decay (See
also~\cite{Frandsen:2010mr}):
\begin{eqnarray}\label{ACRDM}
X\ra e^-\mu^+,\quad e^-\tau^+,\quad \mu^-\tau^+.
\end{eqnarray}
Two comments are in orders. First, if this asymmetry is generated by
asymmetric DM decay, the case in this work, generically $r^+(E)$ has
an energy dependence. Second, in Ref.~\cite{Chang:2011xn} the case
of a fermionic DM decaying into a charged lepton plus a charged
Higgs boson or $W$ boson is considered, but it produces many
anti-protons and thus is disfavored. Moreover, although not shown
explicitly, the fits in this case becomes much worse.
\begin{figure}
\includegraphics[width=80mm,angle=0]{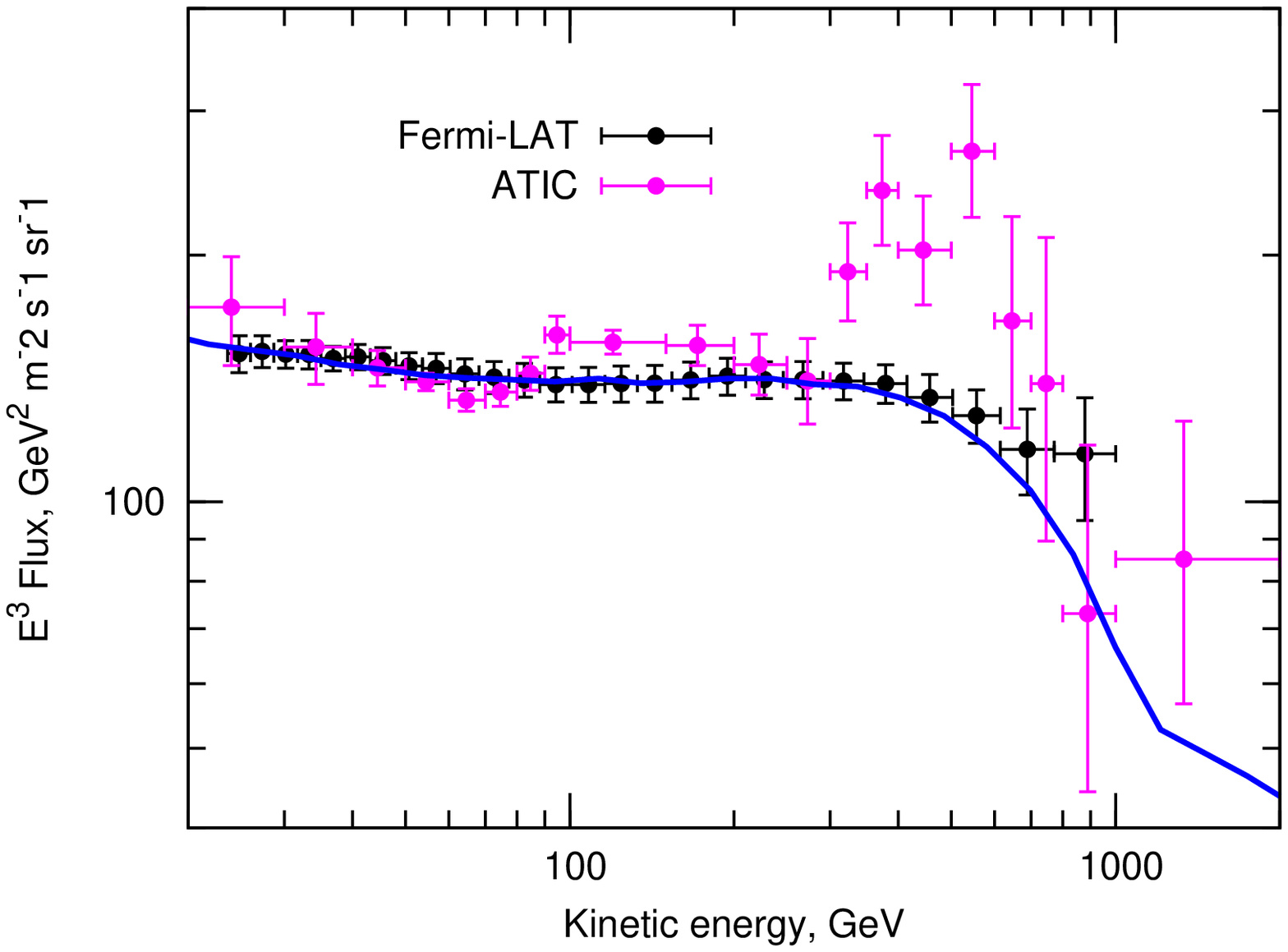}
\includegraphics[width=80mm,angle=0]{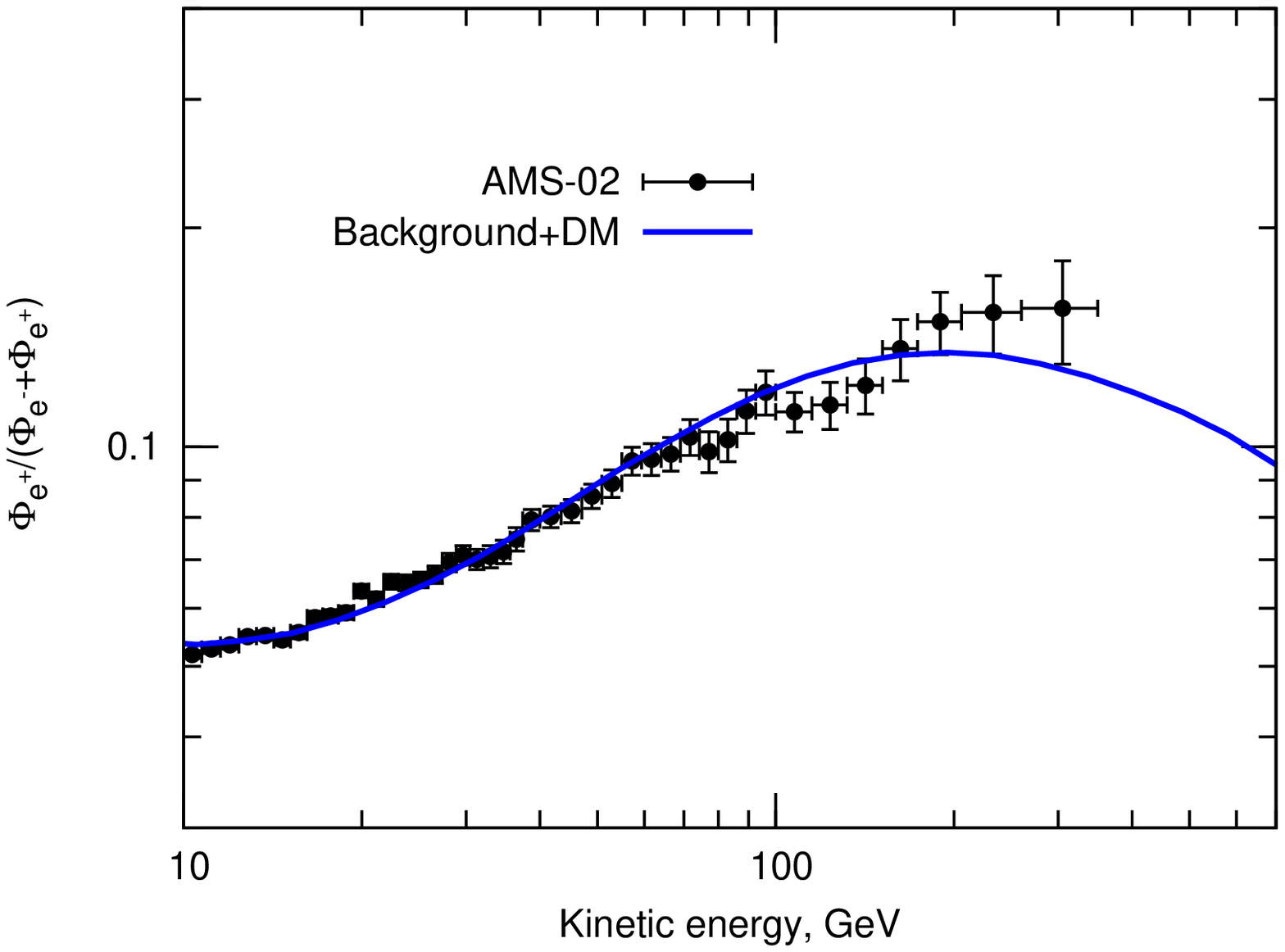}
\includegraphics[width=80mm,angle=0]{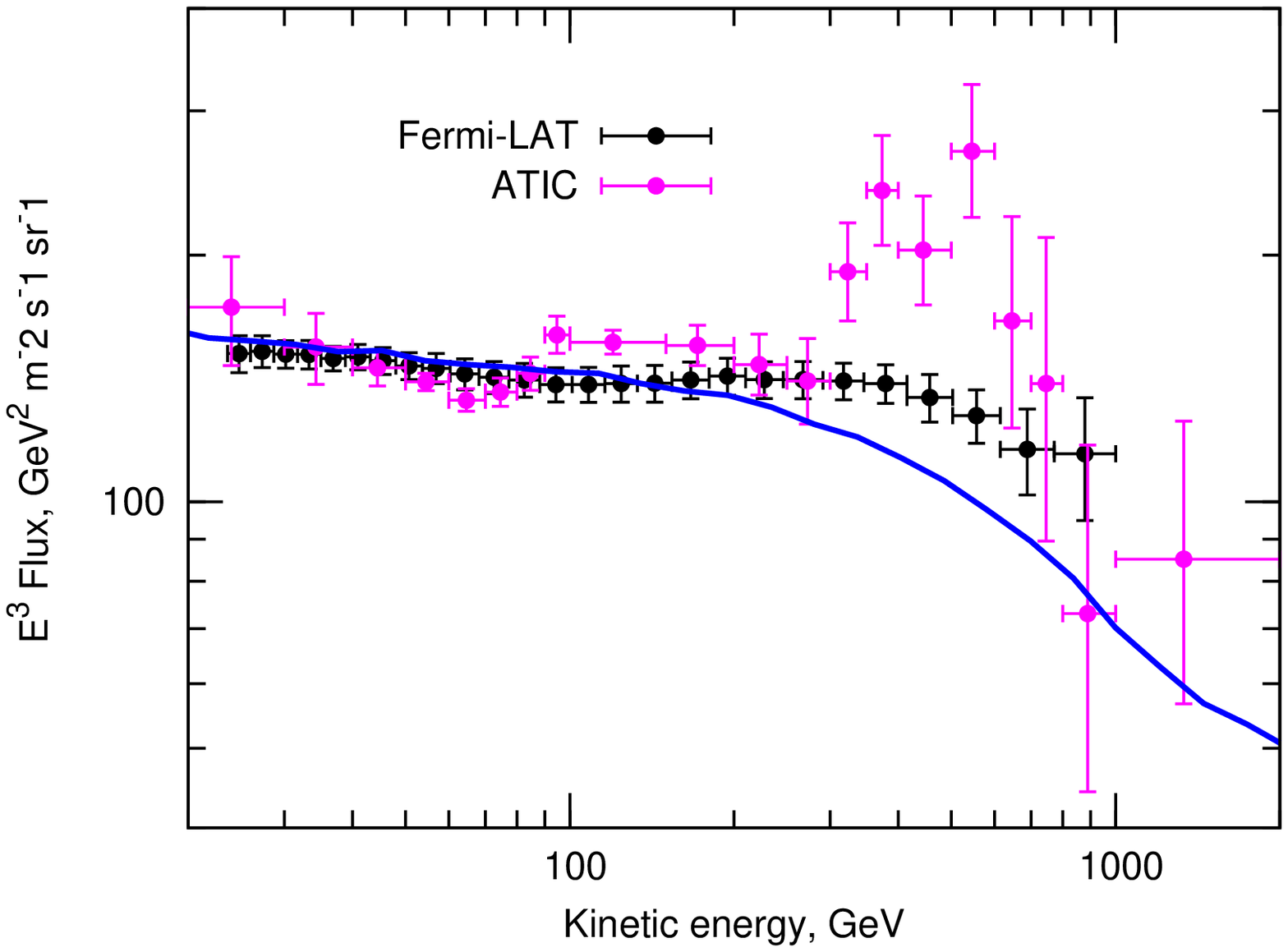}
\includegraphics[width=80mm,angle=0]{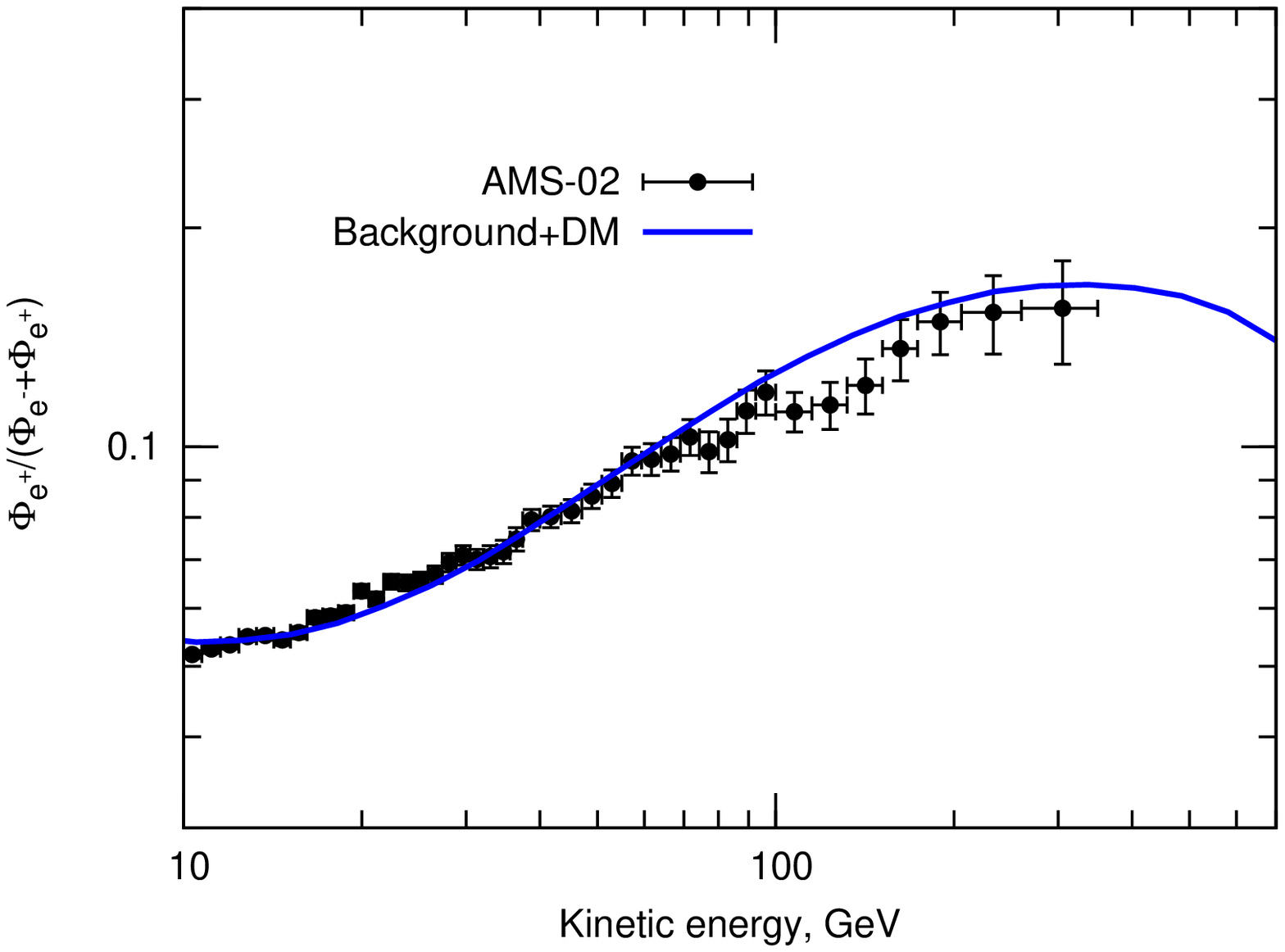}
  \caption{Top-panels: Fits for the AMS-02 positron fraction spectrum
  (right) and the Fermi-LAT $e^{-}+e^{+}$ flux spectrum
  (left), using an asymmetric cosmic ray originating
  in an asymmetric dark matter decaying to $\mu^{-}+\tau^{+}$.
  The best fit is achieved for the dark matter with mass 2.39 TeV
   and lifetime $2.23\times10^{26}~{s}$. Bottom-panels: A comparison to
   the best fit using the symmetric cosmic ray from DM decaying to $\tau^+\tau^-$.}
  \label{fig:1}
\end{figure}

Before heading towards data fitting, we introduce the astrophysical
conventions/notations used in this paper.
The propagation of CR in the Milk Way is
described by a Boltzmann equation. For a given component of CR, we
can obtain its flux by solving this equation using the corresponding
boundary condition. However, owing to the complicated distributions
of sources, interstellar matter, radiation field and magnetic field,
it is difficult to solve it analytically. Thus here we turn to the
numerical code, the GALPROP package~\cite{galprop}, and we modify
the code to include the contribution from DM decay. In the fitting
process, we take the following parameters for the CR propagation:
The diffusion coefficient $D_0=5.3 \times 10^{28}~{\rm
cm^{2}~s^{-1}}$, the diffusion index $\delta=0.33$, the Alfven
velocity $V_{\rm A}=33.5~{\rm km~s^{-1}}$ and the Halo height
$z_{\rm h}=4$ kpc. Furthermore, the injection indexes of nucleon
below and above the break rigidity $\rho_{\rm br}=11.5$ GV are
$1.88$ and $2.39$, respectively. Finally, as for the DM distribution
function, we use the Einasto density profile~\cite{navarro}:
\begin{equation}
\rho(r)=\rho_{s}\exp\left(-\frac{2}{\alpha_s}\left[\left(\frac{r}
{r_{s}}\right)^{\alpha_s}-1\right]\right),
\end{equation}
where $\alpha_s=0.17$, $\rho_{s}\approx 0.14$ GeV cm$^{-3}$ and
$r_{s}\approx 15.7$ kpc.

With our code based on the CosmoMC package~\cite{cosmomc}, we now
perform the Markov Chain Monte Carlo (MCMC) global fitting to
determine the relevant parameters. We find that, among the
three modes listed in Eq.~(\ref{ACRDM}), the best mode is
$X\ra \mu^-\tau^+$, with a $\chi^2=60.26$ and the degrees of freedom
$63$. The corresponding ADM mass and lifetime are 2.39 TeV
and $2.23\times 10^{26}$ s, respectively. The fits for the AMS-02 and
Fermi-LAT data are shown in the top panels of Fig.~\ref{fig:1}. As a
comparison, we also display the fits in the symmetric decaying DM
case, see the bottom panels of Fig.~\ref{fig:1}. The best fit is
achieved for a similar DM, with mass and lifetime 2.70 TeV
and $1.93\times 10^{26}$ s, respectively, but decaying into
$\tau^+\tau^-$. The chi square is much worse, $\chi^2=142.45$
with the degrees of freedom $63$.

\section{Asymmetric cosmic ray from asymmetric DM Decay}

In particle physics, it is natural to expect that the asymmetric CR
originates from asymmetric dark matter (ADM)
decay~\cite{Chang:2011xn,Frandsen:2010mr}. However, for the sake of
naturally solving the cosmic coincidence puzzle, namely
$\Omega_bh^2:\Omega_{\rm DM}h^2\simeq6:1$, the ADM mass is expected
to be around 10 GeV~\cite{Chang:2011xn} rather than at the TeV scale
inspired by solving the AMS-Fermi tension. Additionally, it is
difficult to embed ADM into the simple supersymmetric standard
models (SSMs) such as the minimal-SSM (MSSM) and its singlet (e.g.,
the NMSSM) or right-handed neutrinos extensions, owing to the robust
neutralino mediated charge wash-out effect~\cite{Kang:2011ny}. In
this section we will first construct a simple and natural model to
address these problems and then discuss the phenomenological aspects
of the model.

\subsection{TeV-scale decaying ADM}

Within the conventional ADM framework~\cite{Barr:1990ca}, dark
matter carries a generalized lepton or baryon number through proper
couplings to the standard model (SM) matters. Effectively, this can
be described by the operators such as ${\cal O}_T={\cal O}_{\rm
DM}{\cal O}_{\rm SM}(\ell,q)$, proposed by Ref.~\cite{Kaplan}. Above
some temperature $T_D$, they establish the chemical equilibrium
between the dark sector and visible sector and lead to~\cite{Turner}
\begin{eqnarray}\label{DVeq}
\sum_i\mu_{\phi_i}=0,
\end{eqnarray}
with $\mu_{\phi_i}$ the chemical potentials~\cite{Turner} of the
particles appearing in the previous operators. In other words, they
transfer the visible sector matter asymmetry (assumed to be
generated via some visible sector dynamics) to the hidden sector,
which is symmetric if ${\cal O}_T$ is turned off. Eq.~(\ref{DVeq})
indicates that the chemical potentials of two sectors should be at
the same order.

Further, for the particles in the thermal bath with temperature $T$,
their asymmetries can be expressed with the corresponding chemical
potentials~\cite{Universe¡±}
\begin{eqnarray}\label{npm}
n_+-n_-&=&g\f{T^3}{\pi^2}\f{\mu}{T}\int_0^\infty dx
\f{x^2\exp[-\sqrt{x^2+(m/T)^2}]}{\L\theta+\exp[-\sqrt{x^2+(m/T)^2}]\R^2} \nonumber\\
&\equiv& \left\{ \begin{array}{l}
f_{b}(m/T)\times g_b\f{T^3}{6}\L\f{\mu}{T}\R,\quad {\rm(for\,\, bosons)} \\
f_{f}(m/T)\times g_f\f{T^3}{6}\L\f{\mu}{T}\R,\quad {\rm
(for\,\,fermions)}
\end{array}\right.
\end{eqnarray}
where $\theta$ takes 1 and -1 for a boson and fermion respectively.
$g$ denotes the internal degrees of freedom. The Boltzmann
suppression factors $f_{b,f}(m/T)$ indicate the threshold effects
for heavy particles. For particles in the ultra-relativistic limit,
i.e., $m\ll T$, $f_{b,f}$ tend to 2 and 1, respectively. In the
opposite, the asymmetry of the decoupling particle can be greatly suppressed.
Specified to DM and assuming that the symmetric part of its number
density annihilates away, then the relic densities of baryonic and
dark matters take the ratio:
\begin{eqnarray}\label{puzzle}
\f{\Omega_bh^2}{\Omega_{X}h^2}=\f{m_p}{m_{X}}
\f{\sum_q\mu_q}{g_Xf_{X}\mu_X},
\end{eqnarray}
with $m_X$ the DM mass. Thereby, when the transfer ceases at a
temperature far above $m_X$, just the case for the conventional
scenario~\cite{Barr:1990ca}, it is found that $m_X\sim 10$ GeV is
predicted to solve the cosmic coincidence puzzle. By contrast, when
the chemical equilibrium decoupling happens during the period of DM
entering non-relativistic, then the puzzle has to be resolved by a
heavy DM~\cite{Buckley:2010ui,Graesser:2011wi}. In spite of a
sensitive dependence on the ratio $m/T$, ADM in this scenario still
has a dynamical origin for its similar relic density to the baryonic
matter's (Beyond the chemical equilibrium mechanism to generate DM
asymmetry, there are other ways to get a heavy ADM~\cite{AWMIP}.).

To realize a leptonic decaying ADM, it is tempting to consider that
${\cal O}_T$ not only transfers asymmetry but also provides a path
for DM decaying into leptons. It can be achieved by assuming that
${\cal O}_T$ involve leptons only and $X$ develops a vacuum
expectation value (VEV) $v_X$, which breaks the symmetry protecting
DM stabile. However, bare in mind that here the
decaying DM is extremely long lived, so either a very small $v_X$ or
exceedingly suppression from the operator coefficients is
indispensable. The former, in principle is possible given some
complicated dynamics, while the latter is inconsistent with the TeV
scale ADM set up because it requires the operators to decouple below
$m_X$. Then we are led to the scenario where an extra source
slightly violates the symmetry protecting DM stability. A good case
in point is the $R-$parity-violating SUSY, where the tiny $R-$parity
violations are spectator to the usual dark matter dynamics and only
account for DM late decay. In the following subsection, based on the
$R-$parity-violating SUSY we construct the minimal model for
leptonic decaying ADM at the TeV scale.

\subsection{A minimal supersymmetric model}

Now we embed a heavy decaying ADM in SUSY. Due to the neutralino
mediated wash-out effect~\cite{Kang:2011ny}, ADM is difficult to be
accommodated in the popular supersymmetric models. Within the
chemical equilibrium framework,
Ref.~\cite{Kaplan,Buckley:2010ui,Graesser:2011wi} utilized high
dimension operators like ${\cal O}_T=X^2LLE^c$ with a new and
relatively low cut-off scale, to transfer asymmetry. Here we
construct a more economic model, which even does not incur
any new scales in the superpotential.
The model takes a form of (Actually, it
resembles the supersymmetric inverse seesaw model proposed in
Ref.~\cite{Kang:2011wb}, where a light ADM is studied):
\begin{align}\label{}
 W=&\ld_{ijk} L_iL_jE_k^c+ \L y_{iX}L_iH_u X+\ld_X S X\bar X\R\cr
 +&\L\ld
 SH_uH_d+\f{\kappa}{3}S^3\R+W_{\rm MSSM}.
\end{align}
The corresponding soft terms are implied. $L_i$ and $L_j$ are
asymmetric, and thus $\ld_{ijk}=-\ld_{jik}$. We order $i<j$
hereafter. The superpotential is divided into three sectors: the
$R-$parity-violating sector which accounts for the ADM leptonic
decay, the dark sector which transfers asymmetry between (s)leptons
and DM, and the ordinary NMSSM sector, i.e., terms in the second
line, which dynamically generates the low energy scales via the
singlet VEV $\langle S\rangle\equiv v_s$. In this model the dark
sector fields $(X,\bar X)$ respectively carry lepton number $-1$
and $+1$, by virtue of the renormalizable transferring operator
${\cal O}_T=y_{iX}L_iH_uX$ (A single family will be considered
for simplicity and $y_{iX}\ra y_X$ hereafter).
But the lepton number is explicitly violated by the first
sector, so it fails to guarantee the general structure of the model.
We may have to turn to other symmetries, such as the generalized $Z_3$
symmetry of the NMSSM. We will specifically address this problem
somewhere else.

We now discuss several evident features of the model's parameter
space. First, a large $\ld_X$ is favored to annihilate away
the symmetric part of DM number density. But we require the
model to stay  perturbative up to the GUT scale, so $\ld_X\lesssim0.5$
and then a sufficiently large singlet VEV, says $v_s\sim{\cal O} (5)$ TeV,
is necessary to make $M_X=\ld_X v_s\sim{\cal O}(\rm TeV)$. To that end,
a large $\ld A_\ld$ may be required in the NMSSM Higgs sector.
Next, the soft spectrum is
rather heavy because the TeV-scale ADM is also the lightest
sparticle (LSP). As a consequence, the Higgsino and singlino masses
$\mu=\ld v_s$ and $M_{\wt s}=2\kappa v_s$ should be sufficiently
heavy, which means that both $\ld$ and $\kappa$ should take larger
values. Finally, $y_{X}$ is an important parameter, which controls
the chemical equilibrium decoupling temperature.  We will
discuss this point more detailedly later.

\subsubsection{A Sneutrino-like ADM}

In our model, the scalar components of the chiral superfieds $X$ and
$\bar X$ (denoted by the same letters) are odd under the
ordinary $R-$parity. Moreover, the dark states do not conserve a
separated dark sector symmetry. Thus, the ADM candidate actually
is also the LSP of the whole model. That is to say, ADM must be
the lighter complex scalar from the $(X,\bar X^*)$
mixture~\footnote{Within the MSSM, the left-handed sneutrino in
principle can be a leptonic decaying ADM proposed in the text. This
is achieved by suppressing the neutralino mediated charge wash-out
effect via a very heavy neutralino spectrum, says the bino has a mass
around 100 TeV. We leave such an interesting scenario for future
study.}. We now explicitly check the dark sector mass spectrum. In
the first, the Dirac pair approximately has a mass $M_X$, at the TeV
scale. Secondly, $(X,\bar X^*)$ mix with the left-handed sneutrino
$\wt\nu_{L}$ after electro-weak symmetry breaking. In the basis
$(\wt \nu_L,X^*,\bar X)$, they have a mass square matrix
\begin{align}
\label{scalarsquaremass} \mathcal{M}_{S}^2 = \left(
                         \begin{array}{ccc}
                                   m_{\wt L}^2 & y_XA_{y_X} v_u & y_XM_X v_u\\
                                  & M_X^2+m_{X}^2    & M_X A_{\ld_X} \\
                                  &  & M_X^2+m_{\bar X}^2
                         \end{array}  \right).
\end{align}
Since the mixings with $\wt\nu_L$ are suppressed by $y_X v_u$, then
for a heavier $m_{\wt L}^2$ it is justified to first diagonalize the
$(X,\bar X^*)$ subsystem and get two eigenstates:
\begin{align}
X_{1}=\cos\theta_X X+\sin\theta_X \bar X^*,\quad X_{2}=-\sin\theta_X
X+\cos\theta_X \bar X^*,
\end{align}
with $\theta_X$ the mixing angle. The corresponding eigenvalues are
given by
\begin{align}
m_{1,2}^2=M_X^2+\f{m_X^2+m_{\bar X}^2}{2}\mp \sqrt{(m_X^2-m_{\bar
X}^2)^2+4(M_XA_{\ld_X})^2}.
\end{align}
$X_1$, the lighter eigenstate, is the ADM (We will identify
$X_1|vacuum\rangle$ and $X_1^\dagger|vacuum\rangle$ as DM and
anti-DM, respectively). It is noticed that $m_1^2$ can be much
smaller than $M_X^2$ in the presence of a large $A_{\ld_X}$ or
negative soft mass squares.

The mixing between $X_1$ and $\wt\nu_L$ endows ADM with sneutrino
properties and hence allows it to two-body decay into a pair of
leptons along the operator $LLE^c$. In addition to that, the
mixing may matter in the later discussions on ADM, such as the ADM
chemical equilibrium with neutrinos. So, we later convenience we
estimate the mixing angle:
\begin{align} \label{}
C_{X\wt\nu_L}\simeq\f{y_Xv_u}{m_{\wt L}^2-m_1^2}\L
A_{y_X}\cos\theta_{X}-M_X\sin\theta_{X}\R.
\end{align}
For $y_X\sim{\cal O}(0.1)$, it is can be readily arranged to be less
than ${\cal O}(10^{-4})$, say by lowering $\sin\theta_X$ and
moreover taking a relatively small $A_{y_X}$. A small $C_{X\wt\nu_L}$
helps to alleviate the need for exceedingly small $R-$parity violations.

\subsubsection{ADM at the earlier Universe}

At the earlier Universe, the dark sector enters thermal equilibrium
with the plasma via its significant couplings to $S$. But such
interactions conserve the dark number and thus do not transfer
asymmetry from the visible sector to the dark sector. As mentioned
before, transfer proceeds only after the $LH_uX$ term becomes
active. It leads to the neutralino-neutrino-DM interactions
\begin{align}\label{scatter1}
{\cal L}_{X\nu}\supset y_{X} U_{um}X_1 \bar\nu P_R \chi_m+c.c.
\end{align}
We have used $\wt H_u^0=\sum_{m=1}^5U_{um}\chi_m$ with $\chi_m$
denoting the the NMSSM five neutralinos having Majorana masses
$M_{\chi_m}$. There are terms proportional to $C_{X\wt \nu_1}$
contributing to the above interactions, but they are negligible due
to $C_{X\wt \nu_1}\ll y_X$. Interactions in Eq.~(\ref{scatter1})
give rise to the following scattering
\begin{align}\label{scatter}
 X_1 \nu \leftrightarrow X_1^*\bar \nu,
\end{align}
which then maintains the chemical equilibrium between the dark
states and the light species. The chemical potential of DM is
determined to be
\begin{align}\label{}
\mu_X=-\mu_\nu.
\end{align}
$\mu_\nu$ expresses the lepton chemical potential and is related to
the quark chemical potential. It is positive, so we have a negative
$\mu_X$ and then only DM survives.

\begin{figure}[htb]
\begin{center}
\includegraphics[width=3.8in]{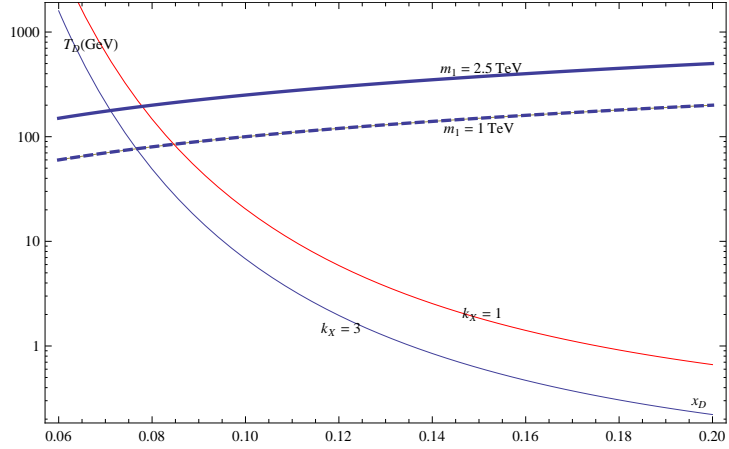}
\end{center}
\caption{\label{XT} A plot of heavy ADM solution to the cosmic
coincidence puzzle, on the $x_D-T_D$ plane. For given ADM mass $m_1$
and the $k_X$ factor, the intersections between the two lines
determine $x_D$ and then the chemical equilibrium decoupling
temperature $T_D$.}
\end{figure}

It is able to precisely determine the ratio $x_D\equiv T_D/m_1$,
with $T_D$ the decoupling temperature at which the chemical
equilibrium of the process Eq.~(\ref{scatter}) breaks. Therefore,
once the ADM mass $m_1$ is fixed, says $m_1=2.39$ TeV in our model
(See section~\ref{fit}), $T_D$ can be determined also. To get that
heavy ADM, we are expecting $x_D\sim{\cal O}(0.1)$, so it is
reasonable to consider $T_D\gtrsim 200$ GeV. It lies above the
sphaleron decoupling temperature, further assuming that it is above
the critical temperature of electroweak phase transition, and then
we have
\begin{align}\label{ADM:model}
\f{n_B}{n_{X_1}}=\f{8}{5k_X}f_X^{-1}(m_1/T_{D}).
\end{align}
To get it we have decoupled the sparticles because we are
considering a rather heavy SUSY spectrum. The factor $k_X$ depends
on the dark sector mass spectrum, and its value varies between 1 and
3. We will comment on it soon later. Now, from Eq.~(\ref{ADM:model})
and Eq.~(\ref{puzzle}) we get $x_D\simeq0.08$ and $T_D\simeq200$ GeV
(This justifies the previous assumption on $T_D$) after taking
$k_X=1$. Note that $x_D$ is not sensitive to $k_X$ and $m_1$, while
$T_D$ has a strong dependence on $m_1$ but not $k_X$, which can be
seen explicitly on Fig.~\ref{XT}. Such features are traced back to
the power law behavior of $f_X^{-1}(x_D)$, see Eq.~(\ref{npm}).

We now proceed to study how $T_D$ arrives. As a rough estimation,
$T_D$ can be determined by equaling $\Gamma_{X\nu}(T_D)$, the
thermal average scattering rate of Eq.~(\ref{scatter}), to the
Hubble expansion rate $H(T_D)=1.66g_*^{1/2}T_D^2/M_{\rm Pl}$. Here
$g_*$ is the effective relativistic degrees of freedom at $T_D$, and
it is at the order of 100. The scattering rate is estimated to be
\begin{align}\label{}
\Gamma_{X\nu}(T_D)\simeq n_X
\langle\sigma_{X\nu}v\rangle\simeq\L0.064g_{X_1}m_1^3x_D^{-3/2}e^{-1/x_D}\R
\f{|y_XU_{um}|^4}{64\pi}\f{M_{\chi_m}^2}{(M_{\chi_m}^2-
m_1^2)^2}2x_D,
\end{align}
with $g_{X_1}=2$ the internal degrees of freedom of the complex
scalar ADM. One of the Feynman diagrams for the scattering cross
section has neutralinos in the $s-$channel, so it shows a resonant
behavior as $M_{\chi_m}$ approach to $m_1$. To decouple the chemical
equilibrium at $T_D\sim200$ GeV, $y_{X}$ can not be too small.
Concretely, from the above equation we need
\begin{align}\label{}
{|y_XU_{um}|^4}\f{M_{\chi_m}^2/m_1^2}{(M_{\chi_m}^2/m_1^2-1)^2}\sim10^{-9}.
\end{align}
Thus, $y_X\sim 0.1$ is needed even if a moderately large resonant
enhancement is provided.

Comments are in orders. First, we have neglected the scattering
process $X_1\nu\ra \wt t t$, of which the scattering cross section
is large and at the order of ${\cal O}(y^2_{iX}y_t^2)$. But stops
are much heavier than $X_1$ while at $T_D$ the typical kinematic
energy available for $X$ and $\nu$ is only $E\sim T_D\sim$ 100 GeV,
which is inadequate to maintain the chemical equilibrium of that
scattering process. Consequently, it decouples at a temperature much
higher than $T_D$ and is irrelevant to our discussion. Next, here
the scattering is just the process which maintains the chemical
equilibrium, and thereby the neutralino mediated wash-out effect is
not of concern. Finally, other dark sector states, the Dirac pair
$(\wt X,\wt {\bar X})$ and $X_2$ store appreciable DM asymmetries
given that their masses are near $m_1$. The former can establish
chemical equilibrium with neutrinos via Higgs mediated scattering.
To account for this uncertainty, we introduce the factor $k_X$ which
has been used previously:
\begin{align}\label{}
k_X=\sum_{X_i}\f{g_{X_i}}{g_{X_1}}\f{f(m_{X_i}/T_D)}{f(m_{1}/T_D)},
\end{align}
with $X_i$ running over $ X_{1,2}$ and the Dirac pair. Clearly, we
have $1\leqslant k_X \leqslant3$, with the upper bound saturated
when $X_i$ are degenerate.

As the Universe cools down to $T_f$, the ADM annihilates away its
symmetric component by means of DM and anti-DM collisions, which
requires a moderately large annihilation rate $\gtrsim 10$
pb~\cite{Graesser:2011wi}. For a TeV scale DM, generically this is
problematic if we require $\ld_X$ and $\ld$ sufficiently small so as
to maintain perturbativity of the model up to the GUT scale.
Fortunately, the singlet sector is heavy (For simplicity, it is
assumed to decouple from the Higgs doublets), and thus the process
$X_1X_1^*\ra H_1H_2$ can be resonantly enhanced by the singlet-like
CP-even Higgs bosons $H_3$ in the $s-$channel. Here $H_{1,2}$ are
$H_u-$ (thus SM-) and $H_d-$like, respectively. In this decoupling
limit, it is ready to calculate the cross section
\begin{align}\label{}
\sigma_{H_1H_2}v\simeq &\f{\ld_X^2}{64\pi}\L\f{M_X^2}{m_1^2}\R
\L\f{(\ld A_\ld)^2}{4m_{1}^4}\R f(4m_1^2/m_{H_3}^2)\cr =&
7.8\times\L\f{\ld_X^2}{0.1}\R\L \f{\ld A_\ld M_X}{10{\rm
\,TeV}^2}\R^2\L\f{2{\rm\, TeV}}{m_1}\R^4\rm pb,
\end{align}
where the Breit-Wigner enhancement factor $f$ is set to 100. Note
that a large trilinear soft term $A_\ld$, which is necessary to get
a large $v_s$, also helps to increase the above cross section. In a
word, it is not difficult to get a sufficiently large cross section
to annihilate away the symmetric part of $n_{X_1}$.

\subsubsection{ADM today}

We now show how does the ADM leptonic decay produce the desired
asymmetric CR in our model. $X_1$ shares the couplings of
left-handed sneutrinos $\wt\nu_{L,i}$, so, along the $R-$parity
violating operators $LLE^c$, it can (only) decay into a pair of
charged leptons. From the data fitting in Section~\ref{fit}, the
$\mu^-\tau^+$ mode is singled out. To this end, we let
$\ld_{123}\,y_{X1}$ dominate over other similar products (This may
be arranged by a flavor symmetry) and then the resulted operator for
ADM decay is
\begin{align}\label{}
{\cal L}_{\rm decay}\supset C_{X\wt\nu_1}\ld_{123}\,X_1\bar \tau  P_L\mu.
\end{align}
Note that its Hermit conjugated term is irrelevant to the ADM decay,
because only DM is left today. In other words, the mode with final
states $\mu^+\tau^-$ is absent in the ADM decay. The $e^-$ and $e^+$
spectra respectively from $\mu^-$ and $\tau^+$ are different, and
therefore we obtain an asymmetric CR. In addition, the ADM life time
is given by
\begin{align}\label{}
\tau(X_1\ra \tau^+\mu^-)\approx 1.7\times 10^{26}\L \f{10^{-4}}{C_{X\wt\nu_L}}\R^2
\L \f{10^{-22}}{\ld_{123}}\R^2\L\f{2\rm \,TeV}{m_{1}}\R s.
\end{align}
When $C_{X\wt\nu_L}\sim10^{-4}$ or even smaller, $\ld_{123}$ can be
obviously larger than those in the previous
works~\cite{Cotta:2010ej}. In the Appendix.~\ref{RPV} we will
present a model to realize the extremely small $\ld_{123}$.

It is worthwhile to note that in our model the sneutrino-like ADM
decay does not produce other significant signatures such as neutrino
flux or anti-protons (But electroweak corrections on the charged
particles or $\tau$ hadronic decay can still induce a correlated
anti-proton signature~\cite{DeSimone:2013fia}). The diffuse gamma
ray from ICS and tau lepton decay is an exception. Actually, as
stressed in the introduction, it is stringently constrained by the
Fermi-LAT gamma ray data even in the decaying DM scenario. To
estimate the surviving status of the ADM with mass 2.39 TeV and
lifetime $2.23\times 10^{26}$s, we refer to several constraints on
the lifetime of DM$\ra\tau^+\tau^-$, using different astrophysical
objects. But the bound here is (roughly) halved because only one
$\tau$ is produced from the ADM decay. Then, the current data of
Fermi-LAT Galactic diffuse emission and galaxy clusters respectively
give the lower bound on the ADM lifetime,
$1.0\times10^{26}$s~\cite{Ackermann} and
$3.0\times10^{25}$s~\cite{Huang}. Thus they do not exclude the
decaying ADM, but the extragalactic gamma ray background, which
gives an lower bound $1.0\times10^{27}$s~\cite{Panci}, does. This
approximate result is consistent with the specific
study~\cite{Masina:2012hg}. So, to avoid this moderate exclusion, we
may have to tune astrophysical parameters such that the fitted
lifetime is allowed.

Finally we briefly discuss the prospect on the direct detection of
the ADM $X_1$. The left-handed sneutrino fraction within the ADM is
negligible, which implies that the $Z-$boson mediated ADM-nucleon
spin-independent scattering is undetectable. Consider the $F-$term
of $S$: $|F_S|^2=|\ld_X X\bar X+\ld H_u^0H_d^0+...|^2$, from which
we get the vertex (The Higgs sector is still in the decoupling
limit)
\begin{align}\label{}
-\f{1}{\sqrt{2}}\ld\ld_X v\sin2\beta\sin2\theta_X|X_1|^2H_1.
\end{align}
The SM-like Higgs boson $H_1=h$ may lead to a detectable
spin-independent scattering cross section, which can be written as
$\sigma_p= f_p^2{\mu_p^2}/{\pi}$~\cite{SUSYDM} with $\mu_p\approx
m_p$ the reduced DM-nucleon mass. Specified to this model, we
have~\cite{Gao:2011ka}
\begin{align} \label{direct:2}
f_p&=-\ld\ld_X\sin2\beta\,\sin2\theta_X\f{m_n}{2m_1}\f{1}{m_h^2}
\L\sum_{q=u,d,s}
{f_{T_q}^{(p)}}+3\times\f{2}{27}{f_{T_G}^{(p)}}\R\cr
&\approx0.8\times10^{-9}\times\L\f{\ld\ld_X}{0.5}\R\L\f{\sin2\beta}{0.4}\R
\L\f{\sin2\theta_X}{1.0}\R\L\f{2{\rm \,TeV}}{m_1}\R {\rm
\,GeV^{-2}},
\end{align}
with $m_h=125$ GeV fixed and values of $f_{T_u}^{(p)}$, etc., given
in Ref.~\cite{Gao:2011ka}. Even for $\sigma_p$ from the above
parameterization to maximize the scattering rate, we have
$\sigma_p\simeq 0.8\times 10^{-10}$ pb. On the other hand, for a TeV
DM the latest XENON100 data~\cite{XENON100} gives the most stringent
upper bound $\sim10^{-8}$ pb, which is far above the estimated
$\sigma_p$. Thus, $X_1$ is hard to be detected, even at the next
around of XENON.


\section{Conclusions and discussions}

The first result of AMS-02 reported the excess of cosmic positron
fraction. This confirms the previous conclusion by PAMELA, but its
spectrum is relatively softer which leads to a tension with the
Fermi-LAT $e^+$ and $e^-$ total flux spectrum measurement. We may
need a somewhat unconventional extra source of CR to resolve the
tension. In this article we propose a solution using the asymmetric
CR from asymmetric dark matter late decay. We find that a multi-TeV
ADM (about 2.4 TeV) asymmetrically decays into $\mu^-\tau^+$ can
indeed significantly improve the fit which adopts the symmetric DM
decay. Based on the NMSSM with $R-$parity-violating SUSY, we
construct a minimal renomalizable supersymmetric model to realize
that scenario. The model only introduces a pair of singlets and
involves no new scales. But the ADM is difficult to be directly
detected.

In the decaying ADM framework to explain the AMS-02 and Fermi-LAT
anomalies, there are some open questions needed to be answered.
First, at the tree-body decay level, it is of interest to consider
modes containing quarks, says $X\ra e^- u\bar d$. After
hadronization, we will have an asymmetric proton spectrum, and $\bar
p/(p+\bar p)<0.5$. Can it lead to an acceptable anti-proton spectrum
by PAMELA? Next, it is not clear whether or not the ADM with multi
decay modes with proper branching ratios can help to further improve
the fits. Last but not the least, distinguishing the asymmetric CR
generated by the ADM mechanism from the one by astrophysical
mechanisms is very important. For example, it may be done by
measuring the $e^-$ spectrum precisely.

To end up this article, we would like to comment on constraints on
the non-annihilating bosonic ADM. From some astrophysical objects
such as the neutron star, it is can be constrained or even excluded
for the light bosonic ADM~\cite{LADM} (but with
exception~\cite{Bell}). However, heavy bosnic ADM has not been
constrained yet~\cite{Kouvaris:2012dz}.

\section*{Acknowledgement}

We thank useful discussions with Yizhong Fan. This research was
supported in part by the China Postdoctoral Science Foundation (No.
2012M521136 and 2013M530006).

\appendix

\section{Generating small $R-$parity violation}\label{RPV}

The decaying ADM life time is extremely long, $\sim10^{26}$s, and
thus how to generate the extremely small $R-$parity violations is
challenging. In this appendix we present a simple model to explain
its smallness. To that end, we introduce a pair of vector-like
lepton doublets $(L_V,\bar L_V)$ with mass around the Planck scale.
They mediate lepton number violations, which are induced by the
$F-$term of some singlet ${\cal X}$ (namely $F_{{\cal X}}\neq0$) to
ordinary leptons. The model is given by
\begin{align}
\f{c_j}{M_{\rm Pl}}\int d^4\theta\bar L_VL_j{\cal X}^\dagger+\int
d^2\theta\ld_{ik}L_iL_V E_k^c+M_VL_V\bar L_V.
\end{align}
Actually, the first term generates a tiny mixing between the
vector-like leptons and the light SM leptons. The mixing is
suppressed by ${F_{{\cal X}}}/{M_VM_{\rm Pl}}$, and as a consequence
the induced coefficients for the $R-$parity-violating operators
$L_iL_jE_k^c$ are estimated as
\begin{align}
\ld_{ijk}\sim\ld_{ik}c_j\f{F_{{\cal X}}}{M_VM_{\rm Pl}}.
\end{align}
It is interesting to note that $\ld_{ijk}$ as small as $10^{-20}$
can be achieved even if $F_{{\cal X}}/M_{\rm Pl}\sim m_{\wt G}$ with
the gravitino mass $m_{\wt G}\sim 1$ TeV, the typical mass scale of
soft terms in the gravity-mediated SUSY-breaking. In other words,
${\cal X}$ may originate in the hidden sector breaking SUSY and,
tampingly, is identified with the hidden sector SUSY-breaking
spurion field.

\end{document}